\documentclass{article}
\usepackage{spconf}
\usepackage{color,slashbox}
\usepackage{math}
\usepackage{enumerate}
\usepackage{theorem}
\usepackage{pifont}
\usepackage{multirow}
\usepackage{hyperref}
\usepackage{cite}
\usepackage{graphicx}
\usepackage{amsmath}
\usepackage{amsfonts}
\usepackage{amssymb}
\usepackage{amsopn,soul}
\usepackage{xspace}
\usepackage{array}
\usepackage{epsfig}
\usepackage{algorithm}
\usepackage{algorithmic}
\usepackage{float,pgfplots}

\DeclareMathOperator*{\argmax}{argmax}
\DeclareMathOperator*{\argmin}{argmin}
\newtheorem{lemma}{Lemma}

\title{Phase Retrieval with a Multivariate Von Mises Prior:\\ from a Bayesian formulation to a lifting solution}
\name{Ang\'elique Dr\'emeau$^{\circledast}$\sthanks{This work has been supported by the DGA/MRIS.}, Antoine Deleforge$^\divideontimes$}
\address{$^{\circledast}$ ENSTA Bretagne and Lab-STICC UMR 6285, Brest, F-29200, France\\$^{\divideontimes}$ INRIA Centre Rennes-Bretagne Atlantique, Campus universitaire de Beaulieu, F-35000 Rennes, France}
\begin{document}
%
\maketitle
\begin{abstract}
In this paper, we investigate a new method for phase recovery when prior information on the missing phases is available. In particular, we propose to take into account this information in a generic fashion by means of a multivariate Von Mises distribution. Building on a Bayesian formulation (a Maximum A Posteriori estimation), we show that the problem can be expressed using a Mahalanobis distance and be solved by a lifting optimization procedure. 
\end{abstract}


\section{Introduction}
Since more than twenty years, phase retrieval has been a constantly filled topic. This is because the problem interests  numerous application domains, from crystallography \cite{Harrison1993} to optical imaging \cite{Dremeau2015a}. Formally, it can be written as follows: given $\yvect\in\mathbb{R}^M$, recover $\xvect\in\mathbb{C}^K$ such as 
\begin{align}
 \yvect = |\Amat\xvect|,
\end{align}
where $\Amat$ is a $M\times K$ known complex-measurement matrix.
Several answers to this non-convex optimization problem have been proposed, that we can roughly divide into three families: \textit{i)} alternating-projection algorithms, where we can find the works of Gerchberg \& Saxton \cite{Gerchberg1972}, Fienup \cite{Fienup1982} or Griffin \& Lim \cite{Griffin1984}, which alternate projections on the span of the measurement matrix and on the object domain, \textit{ii)} algorithms based on convex relaxations, such as the recent \textit{PhaseLift} \cite{candes2013phaselift} and \textit{PhaseCut} \cite{waldspurger2015phase}, which replace the phase recovery problem by relaxed problems that can be efficiently solved by standard optimization procedures, and \textit{iii)} Bayesian approaches, which express the phase recovery problem as the solution of a Bayesian inference problem and apply statistical tools to solve it, such as variational approximations \cite{Schniter2012, Dremeau2015}. 

In the above procedures, the phases are completely missing from the observations: only intensities or amplitudes are acquired. In this paper, we are interested in phase retrieval problems where phases are observed but marred by noise. 
At the interface between the last two above families, we propose a Bayesian formulation of the problem 
and resort to a lifting optimization procedure to solve it. A priori knowledge over observed phases through various probabilistic laws have been exploited in previous works \cite{Gerkmann2014,Colavolpe2005}. Compared to them, our approach presents two appealing novelties: \textit{i)} it is generic in the sense that it can handle multivariate phase priors and thus arbitrary dependencies; \textit{ii)} the proposed Bayesian optimization problem is cast into a generalization of the recently proposed \textit{PhaseCut} problem \cite{waldspurger2015phase}, for which a number of efficient estimation procedures readily exist, including convex relaxations. The last point is made possible by exploiting a previously unseen connection between a multivariate generalization of the Von Mises distribution and the Mahalanobis distance.


\section{Bayesian formulation}
\label{sec:model}

In this section, we introduce the Bayesian modeling that we propose to exploit in the following and discuss its link to the Mahalanobis distance, particularly interesting for the optimization procedure. 

\subsection{Observation model}
Let $M$ sensors record $K$ complex signals through linear instantaneous mixing, in the presence of both additive noise and multiplicative phase noise. The noisy observation $\yvect\in\mathbb{C}^M$ is then expressed as
\begin{equation}
 \yvect = \Diagmat{\phivect}^H\Amat\xvect + \nvect \label{eq:the model}
\end{equation}
where $\Amat\in\mathbb{C}^{M\times K}$ is the mixing matrix, $\xvect\in\mathbb{C}^K$ is the source signal, $\nvect\in\mathbb{C}^M$ is the noise vector, $\phivect=[e^{j\theta_1},\dots,e^{j\theta_{M}}]^\top$
is the phase vector with $\thetavect\triangleq[\theta_1,\dots,\theta_{M}]^\top$ $\in]-\pi,\pi]^{M}$, the operator $\Diagmat{.}$ transforms row- or column-vectors into diagonal matrices and $\cdot^H$ denotes the complex conjugate transpose. For simplicity, we assume that the additive noise is zero-mean \textit{i.i.d.} circular complex Gaussian with variance $\sigma_n^2$.
Note that generalizing subsequent derivations to an arbitrary noise covariance matrix $\Gamma_n$ instead is straightforward with appropriate changes of variable.





\subsection{Von Mises prior}
In the literature, model \eqref{eq:the model} has been already considered in phase retrieval problems with a uniform prior on the phases $\thetavect$ (see \textit{e.g.} \cite{Schniter2012, Dremeau2015}). Here, we look for a more informative model enforcing uncertain structures on and between phases.

Considering phases naturally leads to directional statistics. Among them, the most familiar one is probably the Von-Mises distribution, defined independently for each variable $\theta_m$, $m\in\lbrace 1,\ldots,M\rbrace$ as
\begin{align}
p(\theta_m) = \frac{1}{2\pi I_0(\kappa_m)}\exp\big(\kappa_m \cos(\theta_m-\mu_m)\big), \label{eq:VM}
\end{align}
where $\kappa_m\in\mathbb{R}$ and $\mu_m\in]-\pi,\pi]$ are parameters of the distribution, and $I_0(.)$ is the modified Bessel function of the first kind of order 0. This distribution has been considered in the literature, \textit{e.g.} in \cite{Gerkmann2014}. In practice, it is well-adapted to situations where we want to take into account prior information (such as the mean through parameter $\mu_m$ or the variance through $\kappa_m$) on the phases independently of one another. Its extension to the multivariate case is not straightforward and can take different forms \cite{mardia2008multivariate}. 
In this paper, we assume $\thetavect$ to be distributed according to
\begin{align}
p(\thetavect)=& \frac{1}{C(\kappavect,\Deltamat)}
\exp\big(\kappavect^\top \cvect(\thetavect,\muvect)\nonumber\\
	&- \svect(\thetavect,\muvect)^\top\Deltamat\svect(\thetavect,\muvect)
		- \cvect(\thetavect,\muvect)^\top\Deltamat\cvect(\thetavect,\muvect)\big),\label{eq:MVM}
\end{align}
where $C(\kappavect,\Deltamat)$ is a normalizing constant and functions $\cvect$ and $\svect$ are respectively defined by, $\forall m\in\lbrace 1,\ldots,M\rbrace$,
\begin{align}
 c_m(\thetavect,\muvect) \negmedspace=\negmedspace \cos(\theta_m \negmedspace-\negmedspace \mu_m), \quad
 s_m(\thetavect,\muvect) \negmedspace=\negmedspace \sin(\theta_m \negmedspace-\negmedspace \mu_m).
\end{align}
The matrix $\Deltamat$ is real-symmetric with zeros on its diagonal and captures dependencies between phases. Without loss of generality\footnote{Assuming $\muvect\neq[0,\ldots,0]^\top$ amounts to considering the observation model $\yvect = \Diagmat{\widetilde{\phivect}}^H\widetilde{\Amat}\xvect$ with $\widetilde{\phivect} = \Diagmat{\uvect}^H\phivect$, $\widetilde{\Amat} = \Diagmat{\uvect}\Amat$ and $\uvect=[e^{j\mu_1},\ldots,e^{j\mu_M}]$.}, we will assume in the sequel that $\muvect =[0,\ldots,0]^\top\triangleq \zerovect_M$.
This multivariate extension of the Von-Mises distribution was suggested at the end of \cite{mardia2008multivariate}, but does not seem to have been extensively studied or used. We prefer it here over other alternatives due to the following result (proof in appendix \ref{app:A}):
\begin{lemma}
 \label{LEM:1}
 Let $\muvect=\zerovect_M$, and let $\hat{\thetavect}=[\hat{\theta}_1,\ldots,\hat{\theta}_M]^\top$ maximize the multivariate Von Mises distribution (\ref{eq:MVM}). We have:
 $$
\hat{\phivect}  \triangleq [e^{j\hat{\theta}_1},\dots,e^{j\hat{\theta}_{M}}]^\top = \argmin_{\substack{\phivect \\ |\phi_i|^2=1 \;\forall i}} ||\phivect -\mathbf{1}_M||_{\Gamma_{\phi}}^2
 $$
 where $||\cdot||_{\Gamma_{\phi}}$ denotes the Mahalanobis distance with covariance $\Gamma_{\phi}$, $\mathbf{1}_M\triangleq[1,\ldots,1]^\top$, and $\forall (i,k)\in\lbrace 1,\ldots,M\rbrace^2$
 \begin{align}
  \label{eq:Gammaphi}
  (\Gamma_\phi^{-1})_{ik} = \left\lbrace
  \begin{array}{ll}
   \Delta_{ik} & \text{ if }k\neq i,\\
   \frac{1}{2}\kappa_i - \sum_{l\neq i} \Delta_{il} & \text{ if } k=i.
  \end{array}
  \right.
\end{align}
\end{lemma}
In other words, maximizing the density (\ref{eq:MVM}) can be cast as a quadratically-constrained norm-minimization problem. This type of problem is central in the classical phase retrieval literature (see \cite{waldspurger2015phase}), but does not seem to appear when using other multivariate generalizations of the Von Mises distribution as phase priors, \textit{e.g.}, the one studied in \cite{mardia2008multivariate}.

\section{Phase and Signal Estimation}
\label{sec:solution}
\subsection{Maximum a posteriori}
\label{subsec:map}
Using Lemma \ref{LEM:1}, it follows that the Maximum A Posteriori (MAP) estimate of $\phivect$ within model \eqref{eq:the model} and \eqref{eq:MVM} writes:\\
\begin{align}
\hat{\phivect}_{\text{\tiny{MAP}}}
&=  \argmax_{\substack{\phivect \\ |\phi_i|^2=1 \;\forall i}} \log p(\phivect|\yvect),\\
&=  \argmin_{\substack{\phivect \\ |\phi_i|^2=1 \;\forall i}}\negmedspace \frac{1}{\sigma_n^2}||\yvect\negmedspace-\negmedspace\Diagmat{\phivect}^H\Amat\xvect||^2_2 \nonumber\\
&\quad\quad \quad\quad\quad\quad\quad\quad\quad
+||\phivect-\mathbf{1}_M||_{\Gamma_\phi}^2.\label{eq:MAPphi}
\end{align}
Following a similar idea as in \cite{waldspurger2015phase}, we couple this MAP estimation of the phase vector $\phivect$ with a Maximum Likelihood estimation of the source signal $\xvect$:
\begin{align}
\hat{\xvect}_{\text{\tiny{ML}}} 
& = \argmax_{\xvect} \log p(\yvect;\xvect),\\
& = \argmin_{\xvect} ||\yvect\negmedspace-\negmedspace\Diagmat{\phivect}^H\Amat\xvect||^2_2,\\
& = \Amat^+ \Diagmat{\phivect}\yvect,
\end{align}
where $\Amat^+$ stands for the Moore-Penrose pseudo-inversion of matrix $\Amat$.
Reinjecting this estimate in the MAP problem \eqref{eq:MAPphi} leads to
\begin{align}
\label{eq:phimap}
\hat{\phivect}_{\text{\tiny{MAP}}}\negmedspace
&= \negmedspace \argmin_{\substack{\phivect \\ |\phi_i|^2=1 \;\forall i}}  \frac{1}{\sigma_n^2}||(\Imat_M\negmedspace-\negmedspace\Amat\Amat^+)\Diagmat{\yvect}\phivect||^2_2 \nonumber\\
&\quad\quad \quad\quad\quad\quad+
\left\|
(\Imat_M\quad  -\mathbf{1}_M)
\cdot
\left(
\begin{array}{c}
 \phivect \\
  1
\end{array}
\right)
\right\|_{\Gamma_\phi}^2,
\end{align}
where $\Imat_M$ stands for the identity matrix and the Mahalanobis distance term has been re-written to be homogeneous in $\uvect=[\phivect\quad 1]^\top\in\mathbb{C}^{M+1}$.  Using this trick, it follows that solving \eqref{eq:phimap} is equivalent to solving the $M+1$-dimensional problem
\begin{equation}
\label{eq:uargmin}
\hat{\uvect}= \argmin_{\scriptstyle{\substack{\uvect \\ |u_i|^2=1 \;\forall i}}}  \uvect^H\Qmat\uvect,\; \textrm{where}
\end{equation}
\vspace{-5mm}
\begin{equation}
\Qmat 
=
\begin{pmatrix}
\Mmat + \sigma_n^2\Gamma_\phi^{-1} & -\sigma_n^2\Gamma_\phi^{-1}\mathbf{1}_M\\
-\sigma_n^2\mathbf{1}_M^\top\Gamma_\phi^{-1} & \sigma_n^2\sum_{i,k}(\Gamma_\phi^{-1})_{ik}
\end{pmatrix}\in\mathbb{C}^{(M+1)^2},
\end{equation}
with $\Mmat = \Diagmat{\yvect^H}(\Imat_M\negmedspace-\negmedspace\Amat\Amat^+)\Diagmat{\yvect}$. It is easily verified that if $\hat{\uvect}$ is solution of \eqref{eq:uargmin}, then $\hat{\phivect}_{\text{\tiny{MAP}}} = \hat{\uvect}_{1:M}/\hat{u}_{M+1}$ is solution of \eqref{eq:phimap}.

Interestingly, when 
$\Gamma_\phi^{-1}=\zerovect$ (uninformative prior on phases), \eqref{eq:uargmin} is 
equivalent to the program proposed by Waldspurger et al. \cite{waldspurger2015phase} for classical phase retrieval. They refer to this complex quadratically-constrained quadratic program as \textit{PhaseCut}, in reference to its real counterpart which is known to be equivalent to the classical graph-partition problem \textit{MaxCut} \cite{Goemans1995}. These non-convex problems are NP-hard in general, difficult to solve in practice, and have been extensively studied, yielding a number of efficient optimization schemes for particular instances. The most straightforward approach consists in iteratively minimizing \eqref{eq:uargmin} with respect to each $u_i$ alternatively, which can be done in closed-form \cite{waldspurger2015phase}. Since the problem is non-convex, this 
method is bound to converge to a local minimum which depends on the initialization.

\subsection{Lifting solution}
A particularly popular alternative to solve \eqref{eq:uargmin} is referred to as \textit{Lifting}, and consists in solving the following convex \textit{semi-definite program} (SDP) instead :
\begin{equation}
\label{eq:lifting}
\begin{array}{cc}
\operatorname{argmin}      & \trace \{ \Qmat\Umat \} \\
 \Umat\succeq\zeromat      & \\
 \diagvect{\Umat}=\unmat_M & \\

\end{array}
\end{equation}
where $\succeq\zeromat$ denotes positive semi-definiteness. Note that \eqref{eq:lifting} is a \textit{relaxation} of \eqref{eq:uargmin}, in the sense that if $\hat{\Umat}=\hat{\uvect}\hat{\uvect}^H$ is a rank-1 solution of \eqref{eq:lifting}, then $\hat{\uvect}$ is a solution of \eqref{eq:uargmin}. However, $\hat{\Umat}$ may not always be rank-1 in practice. In the classical prior-less phase retrieval case where $\Gamma_\phi^{-1}\!=\!\zerovect$, 
 the combined extensive research efforts in \cite{candes2013phaselift} and \cite{waldspurger2015phase} lay theoretical grounds providing conditions on $\Amat$ for which solving \eqref{eq:lifting} enables stable recovery of the phase vector $\phivect$ and signal vector $\xvect$ with high probability. Extending these theories to the proposed Bayesian generalization necessitates a deep research investigation, which cannot be tackled within this short paper. Rather, an experimental validation of the lifting approach in the multivariate Von-Mises phase retrieval setting is conducted in Section \ref{sec:exp}.

\subsection{Algorithms}
A large number of efficient generic SDP solvers are available, including interior-point methods \cite{Helmberg1996} or augmented Lagrangian methods \cite{Nesterov2007}. As mentioned in \cite{waldspurger2015phase}, the \textit{block-coordinate descent} (BCD) method proposed in \cite{Wen2012} is particularly simple and efficient for problems of the form \eqref{eq:lifting}, and is therefore used here. In practice, when the obtained solution $\hat{\Umat}$ is not rank-1, a natural approach consists in selecting the leading eigenvector of $\Umat$.

\section{Experiments}
\label{sec:exp}
\pgfplotsset{
ymin=0,
ymax=1,
xlabel={$\sigma_n^2$},
ylabel={$\frac{\vert \hat{\mathbf{x}}^H \mathbf{x} \vert}{\|\hat{\mathbf{x}}\|_2 \|\mathbf{x}\|_2}$},
grid=major,
legend columns=1,
legend style={font=\footnotesize}
}

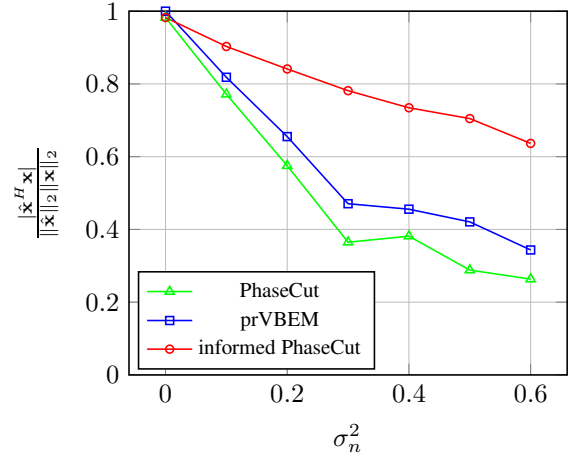
\begin{figure}[t!]
\begin{center}

\begin{tikzpicture}[font=\normalsize]
\begin{axis}
[
mark size = 1.5,
legend style={at={(0.02,0.15)},anchor=west},
line width = 0.6,
scale = 0.85
]
\addplot[color=green,mark size = 2, mark=triangle,solid] table[x index=0, y index=2] {./Corr_PR_noise_variance_dimy256_dims64_k64_ntrial20_varx1_varn0.6_varp11_varpt1_RandMod1_1DVM.dat};
\addlegendentry{PhaseCut}
\addplot[color=blue,mark=square,solid] table[x index=0, y index=4] {./Corr_PR_noise_variance_dimy256_dims64_k64_ntrial20_varx1_varn0.6_varp11_varpt1_RandMod1_1DVM.dat};
\addlegendentry{prVBEM}
\addplot[color=red,mark=o,solid] table[x index=0, y index=3] {./Corr_PR_noise_variance_dimy256_dims64_k64_ntrial20_varx1_varn0.6_varp11_varpt1_RandMod1_1DVM.dat};
\addlegendentry{informed PhaseCut}
\end{axis}
\end{tikzpicture}
\end{center}\vspace{-0.6cm}
\caption{(Averaged) normalized correlation as a function of the variance $\sigma_n^2$ for the \textit{i.i.d.} 1D Von Mises prior.\label{fig:1DVM}}
\end{figure}

In this section, we propose two different experimental setups to assess the relevance of the above procedure. More precisely, we consider two particular cases of the multivariate Von-Mises prior \eqref{eq:MVM} : the 1D Von-Mises distribution and the Markov chain.

For both setups, we confront it to two state-of-the-art phase retrieval algorithms, namely \textit{PhaseCut} \cite{waldspurger2015phase} and \textit{prVBEM} \cite{Dremeau2015}. The first one relies on the same optimization procedure as the one proposed here, but does not exploit any information on the phases to recover. The second one shares the same Bayesian formulation \eqref{eq:the model} as the algorithm proposed here but considers a non-informative, uniform distribution on the phases. In the sequel, we will refer to our approach as ``informed PhaseCut".

We consider the following general experimental setup. Observations are generated according to model \eqref{eq:the model} with $M=256$ and $K=64$. The elements of the dictionary $\Amat$ (resp. vector $\xvect$) are \textit{i.i.d.} realizations of a zero-mean circular Gaussian distribution with variance $M^{-1}$ (resp. $1$). We assess the performance in terms of the reconstruction of the signal $\xvect$. In particular, we consider the correlation between the estimated signal and the one used to generate the data,
$$\frac{\vert \hat{\mathbf{x}}^H \mathbf{x} \vert}{\|\hat{\mathbf{x}}\|_2 \|\mathbf{x}\|_2},$$ as a function of the noise variance $\sigma_n^2$. This figure of merit is evaluated from $50$ trials for each simulation points.

\subsection{1D Von Mises prior}
As a first experimental setup, we consider the case where the phase noise is distributed on each sensor independently of one another according to the Von Mises law \eqref{eq:VM} with parameter $\mu_i = 0$ and $\kappa_i = 1$, $\forall i\in\lbrace 1,\ldots,M\rbrace$.

Figure \ref{fig:1DVM} presents the performance of the three algorithms with this particular prior distribution. As expected, \textit{informed PhaseCut} outperforms the other algorithms, proving a good inclusion of the prior additional information. More particularly, the gap between them increases with the noise variance: for $\sigma_n^2=0.6$, \textit{informed PhaseCut} achieves a correlation around $0.7$ against $0.3$ for \textit{PhaseCut} and \textit{prVBEM}.

\subsection{Markov chain}

\begin{figure}[t!]
\begin{center}

\begin{tikzpicture}[font=\normalsize]
\begin{axis}[
mark size = 1.5,
legend style={at={(0.02,0.15)},anchor=west},
line width = 0.6,
scale = 0.85
]
\addplot[color=green,mark size = 2,mark=triangle,solid] table[x index=0, y index=2] {./
Corr_PR_noise_variance_dimy256_dims64_k64_ntrial20_varx1_varn0.6_varp10.1_varpt0.1_RandMod1_Markov_2.dat};
\addlegendentry{PhaseCut}
\addplot[color=blue,mark=square] table[x index=0, y index=4] {./
Corr_PR_noise_variance_dimy256_dims64_k64_ntrial20_varx1_varn0.6_varp10.1_varpt0.1_RandMod1_Markov_2.dat};
\addlegendentry{prVBEM}
\addplot[color=red,mark=o,solid] table[x index=0, y index=3] {./
Corr_PR_noise_variance_dimy256_dims64_k64_ntrial20_varx1_varn0.6_varp10.1_varpt0.1_RandMod1_Markov_2.dat};
\addlegendentry{informed PhaseCut}
\end{axis}
\end{tikzpicture}

\end{center}\vspace{-0.6cm}
\caption{(Averaged) normalized correlation as a function of the variance $\sigma_n^2$ for the Markov chain prior.\label{fig:markov}}
\end{figure}
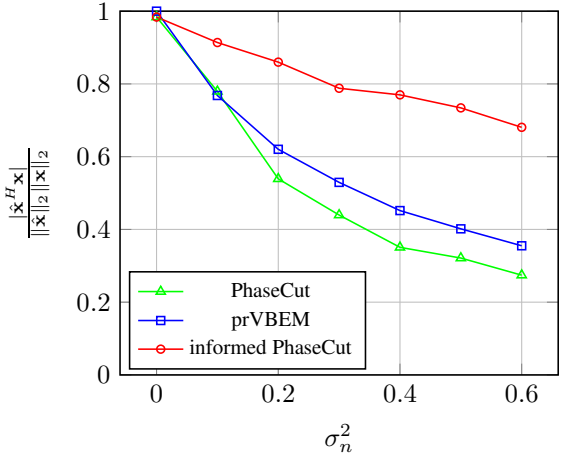
In a second experimental setup, we consider the particular case where only the first two subdiagonals of $\Deltamat$ are non-zero. Considering a small variance of the phases $\thetavect$, straightforward calculus leads to
\begin{align}
&p(\thetavect)
\simeq \frac{1}{C(\kappavect,\Deltamat)}\\
&\exp\left(-\sum_i \left((\Gamma_\phi^{-1})_{ii} \theta_i^2 -2(\Gamma_\phi^{-1})_{i(i-1)}\theta_i\theta_{i-1} +o(\theta_i^2)\right)\right)\negmedspace,\nonumber
\end{align}
where $\Gamma_\phi^{-1}$ is linked to the parameters $\Deltamat, \kappavect$ through \eqref{eq:Gammaphi}. This expression can be directly identified to a Markov chain such as $\forall i\in\lbrace2,\ldots,M\rbrace$,
$\theta_i = a\;\theta_{i-1} + \omega_i,$
where $\omega_i\sim\mathcal{N}(0,\sigma^2_\theta)$, $\forall i$ and $\theta_1\sim\mathcal{N}(0,\sigma^2_\theta)$, providing that
\begin{align}
(\Gamma_\phi^{-1})_{ik} 
= \left\lbrace
\begin{array}{ll}
-\frac{a}{2\sigma_\theta^2} & \text{ if } k= i+1 \text{ or } k= i-1,\\
\frac{1+a^2}{2\sigma_\theta^2} & \text{ if }k= i\neq M,\\
\frac{1}{2\sigma_\theta^2} & \text{ if }k= i= M,\\
0 & \text{elsewhere.}
\end{array}
\right.
\end{align}
We suppose here that $a = 0.8$, $\sigma_\theta^2 =0.1$.

Figure \ref{fig:markov} confirms the good behavior of \textit{informed PhaseCut} observed in the first experiment setup: taking into account the structure of the missing phases, it allows a better estimation (in the sense of the correlation) of the signal of interest $\xvect$. The advantage brought by such prior inclusion increases with the noise variance: 
\textit{informed PhaseCut} reveals here again more robustness. 

\section{Conclusion}
In this paper, we have presented a novel algorithm able to solve the phase recovery problem with a multivariate Von Mises prior distribution. To that end, we have showed that this particular prior information can be efficiently integrated into a Maximum A Posteriori estimation by means of a Mahalanobis distance. The proposed solution relies on a lifting procedure and, to the extent of our experiments, reveals a coherent behavior with regard to non-informed state-of-the-art algorithms.

\appendix

\section{Proof of Lemma \ref{LEM:1}}
\label{app:A}
We have:
\begin{align}
&||\phivect -\mathbf{1}_M||_{\Gamma_{\phi}}^2 \label{eq:prlem}\\
&= (\phivect - \mathbf{1}_M)^H\Gamma_{\phi}^{-1}(\phivect - \mathbf{1}_M)\nonumber\\
& =  \trace \{ \Gamma_{\phi}^{-1}\} +\sum_{i,k} (\Gamma_{\phi}^{-1})_{ik} - 2\sum_i (\Gamma_{\phi}^{-1})_{ii}\cos(\theta_i) \nonumber\\
&\quad- \sum_i\sum_{k\neq i} (\Gamma_{\phi}^{-1})_{ik} e^{-j(\theta_i-\theta_k)}
   - \sum_i\sum_{k\neq i} (\Gamma_{\phi}^{-1})_{ik} e^{-j\theta_i} \nonumber\\
&\quad-\sum_i\sum_{k\neq i} (\Gamma_{\phi}^{-1})_{ki}  e^{j\theta_i},\nonumber\\
& =  \trace \{ \Gamma_{\phi}^{-1}\} +\sum_{i,k} (\Gamma_{\phi}^{-1})_{ik} - 2\sum_i \left(\sum_{k} (\Gamma_{\phi}^{-1})_{ik}\right)\cos(\theta_i)\nonumber\\
&+ \sum_i\sum_{k\neq i} (\Gamma_{\phi}^{-1})_{ik} \cos(\theta_i-\theta_k), \nonumber
\end{align}
where we have assumed that $(\Gamma_{\phi}^{-1})_{ki} = (\Gamma_{\phi}^{-1})_{ik}$, $\forall (i,k)\in\lbrace1,\ldots,M\rbrace^2$, or, in other words, $(\Gamma_{\phi}^{-1})_{ik}\in\mathbb{R}$. Identifying then parameters $\kappavect$ and $\Deltamat$ of the multivariate distribution \eqref{eq:MVM} with \eqref{eq:prlem}, it comes straightforwardly that, under the condition \eqref{eq:Gammaphi},
$$||\phivect -\mathbf{1}_M||_{\Gamma_{\phi}}^2 \propto -\log p(\thetavect)$$ where $\propto$ denotes here
equality up to a constant. 
This means that we can use indifferently the multivariate Von-Mises distribution \eqref{eq:MVM} or the Mahalanobis distance as a 
cost function if we add to the latter the constraint $|\phi_i|=1$, $\forall i \in \lbrace1,\ldots,M\rbrace$.
$\square$

\newpage
\bibliographystyle{IEEEbib}
\small

\bibliography{refs_ICASSP2017_VMphase}

\end{document}